\documentstyle [aps,twocolumn]{revtex}
\input epsf
\tightenlines
\begin{document}
\draft
\preprint{VECC-2000-}
\title{
$J/\psi$ suppression: gluonic dissociation {\em vs.} colour screening}
\author{Binoy Krishna Patra and Dinesh Kumar Srivastava}
\address{Variable Energy Cyclotron Centre, 1/AF Bidhan Nagar, Kolkata 
700 064, India}
\date{\today}
\maketitle

\begin{abstract}
We evaluate the suppression of $J/\psi$ production in an equilibrating
quark gluon plasma for two competing mechanisms: Debye screening of
colour interaction and dissociation due to energetic gluons. Results are
obtained for $S+S$ and $Au+Au$ collisions at RHIC and LHC energies. 
At RHIC energies
the gluonic dissociation of the charmonium is found to be equally
important for both the systems while the screening of the 
interaction plays a significant role  only for the  larger systems.
 At LHC energies the Debye mechanism is found to dominate for both the
systems. While considering the suppression of directly produced 
$\Upsilon$ at LHC energies, we find that only the gluonic dissociation
mechanism comes into play for the initial conditions taken from
the self screened parton cascade model in these studies.
Thus we find that a systematic study of quarkonium suppression for systems of
varying dimensions can help identify the source and  the extent of the
suppression.
\end{abstract}
\pacs{PACS numbers: 12.38M}
\narrowtext

Relativistic heavy ion collision experiments at the CERN SPS are
 believed~\cite{qm99} to
have led to a production of quark gluon plasma - which existed in the early
universe and which may be present in the core of neutron stars.
The last two decades have seen a hectic activity towards identifying unique
signatures of the quark-hadron phase transition. The suppression
of $J/\psi$ production in such collisions has been one of the most
hotly debated signals in this connection. 

The heavy quark pair leading to the $J/\psi$ 
mesons are produced in such collisions on a very short
 time-scale $\sim 1/2m_{c}$,
where $m_c$ is the mass of the charm quark. The pair
develops into the physical resonance over a formation time $\tau_\psi$
and traverses the plasma and  (later)  the  hadronic matter
before leaving the interacting system to decay (into  a dimuon) to 
 be detected.  This long `trek' inside the interacting system is fairly 
`hazardous' for the $J/\psi$.  Even before the resonance is formed
it may be absorbed by the nucleons streaming past it~\cite{GH}. By the
time the resonance is formed,  the screening of the colour forces 
in the plasma may be sufficient to inhibit a binding of the $c\overline{c}$
~\cite{MS,HD}.  Or an energetic gluon~\cite{xu} or a comoving 
hadron~\cite{ramona} 
could dissociate the resonance(s). The extent of absorption
will be decided by a competition between the momentum of the $J/\psi$ and
the rate of expansion and cooling of the plasma, making it
sensitive to such details as the speed of sound~\cite{matsui,dipali}.
Thus a study of $J/\psi$ production  is poised to provide a wealth
of information about the evolution of the plasma and its properties.

It has been shown~\cite{dima} that the nucleonic absorption (the
``normal absorption''), operating on the pre-resonance- which is
yet to evolve into a physical particle- is identical 
for $J/\psi$,  $\psi^\prime$, and  $\chi_c$. This absorption is always
present and is brought about by the nucleons (or the Lorentz-contracted
partonic clouds) streaming past the pre-resonances, as mentioned
earlier. A reliable quantitative estimate
within Glauber model is available~\cite{dima} for this.

In the present work we concentrate on the dissociation of the charmonium
in quark gluon plasma due to colour screening and scattering with
gluons and ask whether we can distinguish between the two mechanisms.
We emphasize that these mechanisms are in addition to nucleonic absorption
mentioned earlier.

In principle the
colour screening is a collective effect, where the presence of a large number
of colour quanta modifies the force between $c$ and $\overline{c}$ so
that, above the critical temperature ($T_c\sim$ 200 MeV), we have:
\begin{equation}
V(r)=-\alpha/r+\sigma r~~\rightarrow~~ V(r)=-\alpha \exp(-\mu_D r)/r
\end{equation}
where $\alpha$ and $\sigma$ (the string tension) are phenomenological
parameters and $\mu_D$ is the Debye mass.

Thus, e.g., the direct production of  the $J/\psi$ is inhibited once 
the Debye mass
is more than 0.7 GeV~\cite{KS}. The gluonic dissociation, on the other hand,
is always possible as long as an energetic gluon can be found. They can always 
be present in the tail of the thermal distributions and thus given sufficient
time, a $J/\psi$ can always be dissociated in a plasma of any temperature! 

Of-course in actual practice the QGP will expand and cool and undergo 
hadronization below the critical temperature $T_c$, and thus the hot medium
will have only a finite life-time. This enriches the competition between the
mechanisms of the gluonic dissociation and the Debye screening for the
charmonium suppression. 
In the present work we show that this also provides us with a handle to
decipher the extent to which each mechanism contributes to the 
suppression of $J/\psi$.

Let us assume that a thermally equilibrated plasma is formed in relativistic
heavy ion collisions at some time $\tau_i$ and that the elastic scattering
among the partons is sufficiently rapid to maintain thermal equilibrium. A 
large number of studies~\cite{klaus,sspc} have indicated that the plasma
 thus produced may not be in  a state of chemical equilibrium and that the
quark and gluon fugacities are less than unity.
 We assume that the chemical equilibration proceeds dominantly
via 
\begin{equation}
gg~\leftrightarrow~ ggg, ~~ gg~\leftrightarrow~q\overline{q}.
\end{equation}
Assuming the evolution to proceed according to Bjorken hydrodynamics, 
the evolution of the parton densities are given by\cite{biro}:
\begin{equation}
\frac{\dot{\lambda_g}}{\lambda_g}+3\frac{\dot{T}}{T}+\frac{1}{\tau}=
R_3(1-\lambda_g)-2R_2\left(1-\frac{\lambda_g^2}{\lambda_q^2}\right),
\end{equation}
\begin{equation}
\frac{\dot{\lambda_q}}{\lambda_q}+3\frac{\dot{T}}{T}+\frac{1}{\tau}=
R_2\frac{a_1}{b_1}\left(\frac{\lambda_g}{\lambda_q}-
\frac{\lambda_q}{\lambda_g}\right),
\end{equation}
\begin{equation}
\left(\lambda_g+\frac{b_2}{a_2}\lambda_q\right)^{3/4}T^3\tau=\mbox{\rm{const}},
\end{equation}
where $a_1=16\zeta(3)/\pi^2 \approx 1.95$, $a_2=8\pi^2/15 \approx 5.26$,
 $b_1=9\zeta(3)N_f/\pi^2 \approx 2.20$, and $b_2=7\pi^2 N_f/20 \approx 6.9$.
 The expressions for the density and velocity weighted reaction rates,
\begin{equation}
R_3=\frac{1}{2}<\sigma_{gg\rightarrow ggg} v > n_g~,
R_2=\frac{1}{2}<\sigma_{gg\rightarrow q\overline{q}} v >n_g
\end{equation}
can be found in Ref.~\cite{biro}. 

The results for the time evolution of the
fugacities and the temperature for the initial conditions obtained 
from the self screened parton cascade model~\cite{sspc} for $Au+Au$ 
collisions at
RHIC and LHC energies are given in Ref.~\cite{munshi}. For the $S+S$ 
collisions we assume that while the initial fugacities are  same as
those for the $Au+Au$ system, the  
initial temperatures are estimated by assuming
that it scales as $T_i\sim A^{\mbox{0.126}}$. This is motivated by
a recent study on the basis of parton saturation~\cite{eskola} 
which also suggests that the initial
number density divided by $T_i^3$ is  nearly independent of the mass-number
 of the nuclei.
This, we believe, provides a useful initial guess, even though the conditions
envisaged for self screening are not strictly met for $S+S$ at RHIC.
For the sake of completeness, we have given the initial conditions in
Table 1. It may be noted that these are different from those used in
Ref.~\cite{xu}, which were `inspired' by the HIJING model and which had,
for example, much smaller fugacities. (We have verified that our computer
program fully reproduced the results of Ref.~\cite{xu}, with the initial
conditions given there.)

We shall also introduce a energy density profile such that,
\begin{equation}
\epsilon(\tau_i,r)=(1+\beta)<\epsilon_i>(1-r^2/R^2)^{\beta}\Theta(R-r)
\end{equation}
where $\beta=1/2$, $R$ is the transverse dimension of the
system and $r$ is the transverse distance, and $<\epsilon_i>$
is the energy density obtained by taking the initial temperature as
$T_i$ and fugacities as $\lambda_i$~\cite{sspc}.  The profile plays an 
important role in defining the boundary of the  hot and dense  deconfined
matter. 

Having obtained the density of the partons we estimate the Debye mass of the
medium as
\begin{equation}
\mu_D^2=\kappa^2 \times 4\pi\alpha_s(\lambda_g+N_f\lambda_q/6)~T^2
\label{mud}
\end{equation}
where we have arbitrarily taken $\kappa$ as 1.5 to account for the
corrections~\cite{kajantie} to the lowest order perturbative QCD 
which provides the above expression for $\kappa=1$. Results for other values
of $\kappa$ are easily obtained. We shall assume that the $J/\psi$
can {\em not} be formed in the region where $\mu_D$ is more than 0.7 GeV.
We can then estimate the survival probability of the directly
produced $J/\psi$ as a function of its transverse momentum $p_T$
by proceeding along the lines of Ref.~\cite{matsui,dipali,KS}.

In order to estimate the gluonic dissociation we recall~\cite{BP} that the short
range properties of the QCD can be used to derive the gluon-$J/\psi$
cross-section as:
\begin{equation}
\sigma(q^0)=\frac{2\pi}{3}\left(\frac{32}{3}\right)^2 \frac{1}
{m_C(\epsilon_0 m_C)^{1/2}}\frac{(q^0/\epsilon_0-1)^{3/2}}{(q^0/\epsilon_0)^5},
\label{sigma}
\end{equation}
where $q^0$ is the gluon energy in the rest-frame of $J/\psi$ and $\epsilon_0$
is the binding energy of the $J/\psi$. The expression for the
thermal average of this cross-section  $<v_{\mbox{\rm{rel}}}\sigma>$ is given 
in Ref.~\cite{xu}. (See, also Ref.~\cite{gypsy} for an interesting alternative
approach.)

We wish to have a quantitative comparison of these two processes and therefore
 it is imperative that we compare their results for similar conditions.
Thus, exactly as while dealing with Debye screening, we assume that the  
$c\overline{c}$ produced intially takes a finite amount of time $\sim 0.89$
fm/$c$ in its rest frame to evolve into the physical resonance. This can 
get large due to time dilation, in the frame of the plasma, leading  to the 
characteristic $p_T$ dependence of the survival probability for the
$J/\psi$ discussed in the literature. 

We argue that the gluon-$J/\psi$ cross-section also attains its 
full value only after the $c\overline{c}$ pair has evolved into the
physical resonance. We assume that this evolution of the cross-section
can be parametrized as
\begin{equation}
\sigma=\left\{ \begin{array}{ll}
\sigma_0\left(\tau/\tau_\psi\right)^\nu  & \mbox{\rm{if} 
                             $\tau \leq \tau_\psi $}\\
\sigma_0  & \mbox{\rm{if} $\tau > \tau_\psi$}
\end{array}
\right.
\end{equation}
similarly to the case when the nuclear absorption is considered~\cite{bl},
where $\sigma_0$ is the cross-section estimated earlier (Eq.\ref{sigma}).
A similar assumption  was invoked by Farrar et al.~\cite{farrar}
when the $Q\overline{Q}$-system evolves as it moves away from the 
point of hard interaction.  One may imagine that 
this amounts to assuming that the effective cross-section
scales as the transverse area of the system relative to the size
it attains when it is fully formed. In the present work
we follow, Blaizot and Ollitrault~\cite{bl} who have 
used $\nu=2$. Farrar et al.~\cite{farrar2} have suggested that
$\nu=1$ corresponds to a quantum diffusion
 of the quarks while $\nu=2$
would correspond to maximal rapid (classical) expansion. Legrand 
et al.~\cite{legrand} have used $\nu=1$ in a recent study.

 This aspect is 
in contrast to the work of Xu et al~\cite{xu} where a fully formed
$J/\psi$ is assumed to exist right at the initial time in the plasma. We shall
see that ignoring the formation time leads to an enhanced suppression of the
charmonium.  

We can now easily estimate the time spent by the $J/\psi$ in the deconfined 
medium for a given $p_T$ and get the survival probability following
Ref.~\cite{xu}.

In Fig.~1 we show our results for RHIC energies for $S+S$ and $Au+Au$
collisions. We see that the combination of a finite formation time
and (reasonably) large $\mu_D$ required to inhibit the formation of
the  directly produced $J/\psi$ in the plasma ensures that the
mechanism of Debye screening is not effective in suppressing its
production. However the gluonic dissociation leads to a suppression
of the $J/\psi$ formation even after the moderating effect of 
the inclusion of formation time is included.

The situation for the larger (and hotter) volume of plasma produced in
$Au+Au$ collisions is much richer in detail.  We see that while the
$J/\psi$s having lower transverse momenta are more strongly suppressed
due to the Debye mechanism, those having higher transverse momenta
are more suppressed by the mechanism of gluonic dissociation. In fact
 we see that while the Debye screening has become quite ineffective 
for $p_T>$6 GeV, the gluonic dissociation continues to be operative.
The different results obtained here compared to authors of Ref.~\cite{xu}
(when the formation time considerations are ignored) are solely due to
the SSPC initial conditions (Table 1) used here.

The corresponding results at LHC energies are shown in Fig.~2. Now we
see that the Debye screening is more effective in suppressing the
production of the directly produced $J/\psi$ at all the momenta considered,
provided we include the considerations of the formation time while
evaluating the gluonic dissociation, for both the systems. 

Of-course in a model calculation one can arbitrarily enhance the 
impact of Debye screening by taking a larger  value for the coefficient
 $\kappa$ (Eq.\ref{mud}). This sensitivity would be useful for determining
its precise value~\cite{ramona-up}.

The treatment outlined here can be extended to the case of
$\Upsilon$ production studied in great detail by  the authors of 
Ref.~\cite{dipali,ramona-up}, for example. We give the results only for
the LHC energies, for the directly produced $\Upsilon$ (Fig.~3).
We find that both for the light as well as the heavy systems the
Debye mechanism is not at all able to inhibit the formation of the
directly produced $\Upsilon$s, though the gluonic dissociation
leads to a considerable suppression, with the changes brought 
about by the inclusion of the formation time seen earlier
for the $J/\psi$s. This is easily seen to be the consequence of
the initial conditions used here, which have chemically non-equilibrated
plasma leading to small Debye mass, even though the temperature
is rather high. By the time  the $\Upsilon$ is formed the
Debye mass drops below the value of $\sim$~ 1.6 GeV, required
to inhibit its formation, causing it to escape unscathed.

Before summarizing, let us discuss some of the assumptions made in this
work. We have, so far considered only the dissociation of the directly
produced $J/\psi$. Of-course, it is well known that up to 30\% of the 
$J/\psi$s seen in these studies may be produced from a decay of $\chi_c$
and up to 10\% or so may come from the decay of $\psi^\prime$, which however
is also easily dissociated by a moderately hot (confined) hadronic
matter and is unable to play a decisive role in distinguishing
confined matter from a deconfined matter. In order to include the effect
of these resonances, we should also have access to $g-\chi_c$ cross-section.
This would involve extending the method of Ref.~\cite{khar} to 
the case of charmonium in $1p$ state. However this is not quite
easy as the assumption $\epsilon_0 \gg \Lambda_{\mbox{QCD}}$
 used in the above reference are not strictly 
valid for this case, as the binding energy of $\chi_c$ is only about 240 MeV. 

Still, from the considerations of binding energy alone, one may expect the
$g-\chi_c$ cross-sections to be larger. However, the Debye mass required to
inhibit the formation of $\chi_c$ is also smaller and thus this competition
between the two mechanisms will continue. The inclusion of the transverse
expansion and the possibility of a different speed of sound than the value
of $1/\sqrt{3}$ assumed here will also add to the richness of the information
likely to be available from such studies. Of-course a full study will
additionally include the effect of the nuclear and the co-mover absorption,
before these interesting details are investigated.

The incorporation of the formation time is interesting for one more reason.
The pre-equilibrium stage (before the time $\tau_i$) may be marked by  
presence of gluons of high transverse momenta, as a result of first
hard collisions, and one may imagine that they play an important role
in suppression of charmonium formation. This is unlikely for two 
reasons. Firstly, the gluon-charmonium cross-section  drops rapidly
as the gluon momentum increases~\cite{khar} after reaching a
peak around $p\sim$ 1 GeV. Secondly we expect
 these cross-sections to be further suppressed during the formation era
due to the considerations of the formation time. 

While considering the suppression of $\Upsilon$, we found that only the
mechanism of gluonic dissociation is playing a role. This happens as
the initial conditions used here involve a chemically non-equilibrated
plasma. If the initial fugacities were to be larger, the Debye
screening would also play a role, which will definitely be a
good check on these.

In brief, we have seen that while the gluonic dissociation of the $J/\psi$
is always possible, the Debye screening is not effective in the case
of small systems at RHIC energies. For the larger systems, the Debye screening
is more effective for lower transverse momenta, while the gluonic
dissociation dominates for larger transverse momenta. At LHC energies
the Debye screening is the dominant mechanism of $J/\psi$ suppression
for all the cases and momenta studied. We have also seen  that the inclusion
of the formation time of the $J/\psi$ plays an interesting role in
reducing the role of the gluonic dissociation. As an interesting result,
we find the gluonic dissociation to be substantial but the Debye screening to
be ineffective for $\Upsilon$ suppression at the LHC energy. 
This may of-course change if different initial conditions
and screening criteria~\cite{dipali,ramona-up} are employed.
 
\section*{Acknowledgements} We thank Dr. Dipali Pal for collaboration during
the early phases of this work and Prof. Helmut Satz for useful comments.
We also thank Prof. Joseph Kapusta for useful correspondence.

\newpage
\begin{table}[h]
\caption{Initial values for the time, temperature,  
           fugacities etc. for Au+Au~\protect\cite{sspc}
and S+S at RHIC and LHC.}
\begin{tabular}{lcc}
$Au+Au$& RHIC & LHC \\
\tableline
   &  & \\
$\epsilon_i$ (GeV/fm$^3$) & 61.40 & 425 \\
   &  & \\
$T$  (GeV) & 0.668 & 1.02 \\
  &  & \\
$\tau_0$ (fm) & 0.25 & 0.25 \\
  & & \\
$\lambda_g$ & 0.34 & 0.43  \\
  &  &\\
$\lambda_q$ & 0.068 & 0.086 \\
  & &\\
\tableline
   &  &  \\ 
$S+S$& RHIC & LHC \\
   &   & \\
\tableline
   &  & \\
$\epsilon_i$ (GeV/fm$^3$) & 24.3 & 170 \\
  &  &  \\
$T$  (GeV) & 0.531 & 0.811 \\
  &  &  \\
$\tau_0$ (fm) & 0.25 & 0.25 \\
  &  &  \\
$\lambda_g$ & 0.34 & 0.43  \\
  &  &\\
$\lambda_q$ & 0.068 & 0.086 \\
  & &\\
\end{tabular}
\end{table}
\begin{figure}
\epsfxsize=3.25in
\epsfbox{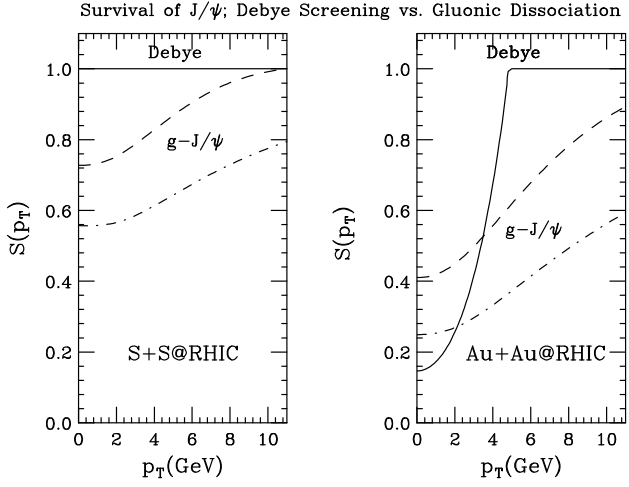}
\vskip 0.2cm
\caption{ Survival probability of directly produced $J/\psi$ at RHIC
energies due to screening of colour interaction (solid curve)
and gluonic dissociation in quark gluon plasma. The dashed curve
gives the latter with inclusion of formation time of the charmonium
while the dot-dashed curve gives the same with the assumption that 
a fully formed $J/\psi$ is available at $\tau=\tau_i$ when the 
plasma is formed.
}
\end{figure}

\begin{figure}
\epsfxsize=3.25in
\epsfbox{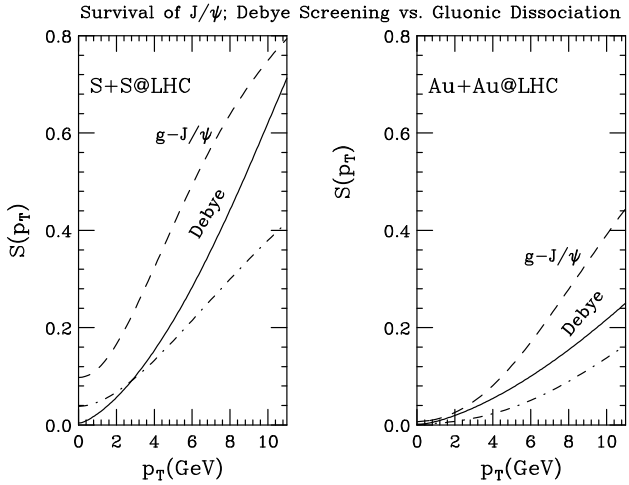}
\vskip 0.2cm
\caption{
Same as Fig.~1 at LHC energies.
}
\end{figure}

\begin{figure}
\epsfxsize=3.25in
\epsfbox{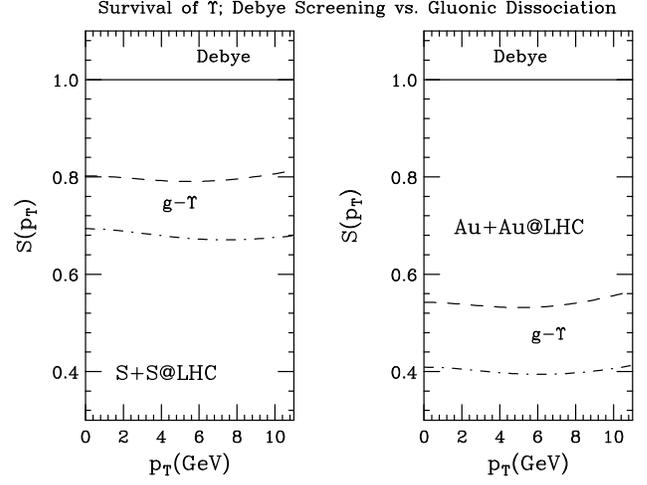}
\vskip 0.2cm
\caption{
Same as Fig.~1  for $\Upsilon$ at LHC energies. The Debye screening
is absent for the initial conditions~\protect\cite{sspc} used here.
}
\end{figure}

\begin{references}

\bibitem{qm99} See e.g.,
M. C. Abreu et al., NA50 Coll. Phys. Lett. B 477 (2000)28;
{\sl Proc. Quark Matter '99}, Nucl. Phys. A  661 (1999);
U. Heinz and M. Jacob, e-Print nucl-th/0002042;
 {\sl Physics and Astrophysics of Quark Gluon Plasma, (Proc. ICPA'97),
 Ed. B. C. Sinha, D. K. Srivastava, and Y. P. Viyogi}, (Narosa Publishing
 House, New Delhi, 1998).

\bibitem{GH} C. Gerschel and J. H\"{u}fner, Phys. Lett. B 207 (1988) 253.

\bibitem{MS} T. Matsui and H. Satz, Phys. Lett. B 178 (1986) 416.

\bibitem{HD} H. Satz and D. K. Srivastava, Phys. Lett. B 475 (2000) 225.

\bibitem{xu} X.-M. Xu, D. Kharzeev, H. Satz, and X.-N. Wang, Phys. Rev.
C 53 (1996) 3051.

\bibitem{ramona} See e.g., R. Vogt, Phys. Rep. 310 (1999) 197;

\bibitem{matsui} M.-C. Chu and T. Matsui, Phys. Rev. D 37 (1988) 1851.

\bibitem{dipali} D. Pal, B. K. Patra, and D. K. Srivastava, Eur. Phys. Jour.
C 17 (2000) 179.

\bibitem{dima} D. Kharzeev and H. Satz, Phys. Lett B 366 (1996) 316.

\bibitem{KS} J. P. Blaizot and J. Y. Ollitrault, Phys. Lett. B 199
(1987) 499; F. Karsch and H. Satz, Z. Phys. C51 (1991) 209.

\bibitem{klaus} K. Geiger, Phys. Rep. 258 (1995) 237.

\bibitem{sspc} K. J. Eskola, B. M\"{u}ller, and X.-N. Wang, Phys. Lett.
B 374 (1996) 20.

\bibitem{biro} T. S. Biro, E. van Doorn, M. H. Thoma, B. M\"{u}ller, and
X.-N. Wang, Phys. Rev. C 48 (1993) 1275.

\bibitem{munshi} D. K. Srivastava, M. G. Mustafa, and B. M\"{u}ller, 
Phys. Rev. C 56 (1997) 1064.

\bibitem{eskola} K. J. Eskola, K. Kajantie, P. V. Ruuskanen, and K.
Tuominen, Nucl. Phys. B 570 (2000) 379.

\bibitem{kajantie} K. Kajantie et al., Phys. Rev. Lett. 17 (1997) 3130.

\bibitem{BP} M. E. Peskin, Nucl. Phys. B 156 (1979) 365;
 G. Bhanot and M. E. Peskin, Nucl. Phys. B 156 (1979) 391.
(See also, D. Kharzeev and H. Satz, Phys. Lett. B 344 (1994) 155 for
a recent application of these results.)

\bibitem{gypsy} S. C. Benzahra, Phys. Rev. C 61 (2000) 064906.

\bibitem{bl} J. P. Blaizot and J. Y. Ollitrault, in {\em Quark Gluon Plasma}
Ed. R. C. Hwa, World Scientific, Singapore, p. 531.

\bibitem{farrar} G. R. Farrar, H. Liu, L. L. Frankfurt, and M. I.
Strikman, Phys. Rev. Lett. 61 (1988) 686.

\bibitem{farrar2} G. R. Farrar, L. L. Frankfurt, and M. I.
Strikman and H. Liu, Phys. Rev. Lett. 64 (1990) 2996.

\bibitem{legrand}  L. Gerland,  L. Frankfurt, and M. I.
Strikman, H. St\"ocker, and W. Greiner, Nucl. Phys. A 663 (2000) 1019.

\bibitem{ramona-up} J. Guinon and R. Vogt, Nucl. Phys. B 492 (1997) 301.

\bibitem{khar} D. Kharzeev and H. Satz, in {\em Quark Gluon Plasma 2}
Ed. R. C. Hwa, World Scientific, Singapore, 1995, p.395.

\end{references}
\end{document}